\documentclass[%
amsmath,amssymb,
preprint,%
]{revtex4-1}

\usepackage{graphicx}
\usepackage{dcolumn}
\usepackage{bm}

\usepackage[utf8]{inputenc}
\usepackage[T1]{fontenc}
\usepackage{mathptmx}
\usepackage{bbold}
\usepackage{tabularx}
\usepackage{xcolor}
\usepackage{enumitem}
\usepackage{hyperref}

\begin{document}
	
	
	\title[Sample title]{Probing the acoustic losses of graphene with a low-loss quartz bulk-acoustic-wave resonator at cryogenic temperatures}
	
	%
	\author{Serge Galliou}
	\email{serge.galliou@femto-st.fr}
	\author{J\'{e}r\'{e}my Bon}
	\altaffiliation[Now at ]{Lab. Mat\'{e}riaux et Ph\'{e}nomènes Quantiques, CNRS UMR 7162, Univ. Paris Diderot, 75013 Paris, France.}
	\author{Philippe Abb\'{e}}
	\author{R\'{e}my Vicarini}
	\affiliation{FEMTO-ST, Univ. Bourgogne Franche-Comt\'{e}, CNRS, ENSMM 26 rue de l'\'{E}pitaphe, 25030 Besan\c{c}on, France}
	
	\author{Michael E. Tobar}
	\email{michael.tobar@uwa.edu.au}
	\author{Maxim Goryachev}
	\email{maxim.goryachev@uwa.edu.au}
	\affiliation{ARC Centre of Excellence for Engineered Quantum Systems, Department of Physics, The University of Western Australia, 35 Stirling Hwy, M013, Crawley 6009, WA, Australia}

	
	\begin{abstract}
		We observe mechanical effects of an exfoliated graphene monolayer deposited on a quartz crystal substrate designed to operate as an extremely low-loss bulk-acoustic-wave cavity at liquid-helium temperature. 
		This is achieved by sensing overtones of the three thickness eigen-modes of the so-called SC-cut, since all three modes, two shear mode and one extensional mode, can be electrically probed with such a crystal cut.
		From quality-factor measurements, the mechanical losses of the adhesive graphene monolayer are assessed to be about $8 \times 10^{-4}$ at $4$~K in the best case. They are therefore significantly greater than those already reported for suspended membranes but also for adherent layers on $SiO_2/Si$ substrates operating in torsional modes. 
		In fact, results reveal that surface scattering occurs due to a roughness degradation of a factor 7. 
		In addition, the mechanical losses presented here are also placed in the context of a device submitted to thermomechanical stresses, but which are not the only ones existing. Some of them could be intrinsic ones related to the deposition process of the graphene layer. Based on a force-frequency theory applied to the three thickness modes which react differently to stresses, it is demonstrated that this stress effect actually entangled with that of mass loading reconciles the experimental results.
	\end{abstract}
	
	\maketitle
	

%
\section{Introduction}
\label{sec:intro}
Bulk acoustic wave (BAW) devices are widely used in research and industry as resonators/cavities, filters or sensors, including Quartz Crystal Microbalance (QCM)~\cite{schedin2007, Quang2014}, for a large variety of applications. Beyond these usual applications at room temperature, it has been demonstrated that plano-convex BAW cavities made of premium-quality quartz and designed to trap the acoustic energy can exhibit Quality factors greater than a billion in the frequency range $1-200$~MHz at liquid helium temperature~\cite{GalliouAPL, galliouScRip2013} when packaged like the device used in this study. In these conditions they become very attractive for various experiments in fundamental physics~\cite{Gory14,Muller2016,Goryachev18} as well as for hybrid quantum systems~\cite{Kotler}, optomechanics~\cite{Aspelmeyer2014, Carvalho2019, Rakich2018}, etc.
With that in mind, the BAW device described in this paper has already been operated as an optical cavity~\cite{BonJAP, Kevin2021}. Consequently, the ability of these devices to be simultaneously both an acoustic and an optical cavity makes them a natural candidate for optomechanical experiments. Although, the material based interaction strength between optical and acoustic fields within the same volume of the cavity remains low. In addition to coupling in bulk, one can be enhanced on a boundary by depositing a mirror coating. On the other hand, it has been demonstrated~\cite{RSIcoating, GalliouTUFFC2016} that deposition of traditional metallic coatings like chromium and gold leads to significant degradation of acoustic quality factors. This motivates investigations of effects of ultra thin graphene layers on BAW devices with a promise of minimizing the loading impact on mechanical losses (i.e. without reducing $Q$-factors)~\cite{Qian2015, Knapp2018}. Even if the device described in the present work is too much complicate to be disseminated as a sensor solution, results experienced from it under unusual conditions are still relevant for sensing applications and deserve to be shared.

\section{Materials and Methods}
\subsection{The quartz crystal acoustic cavity}
\begin{figure}[h!]
	\centering
	\includegraphics[width=2.25in]{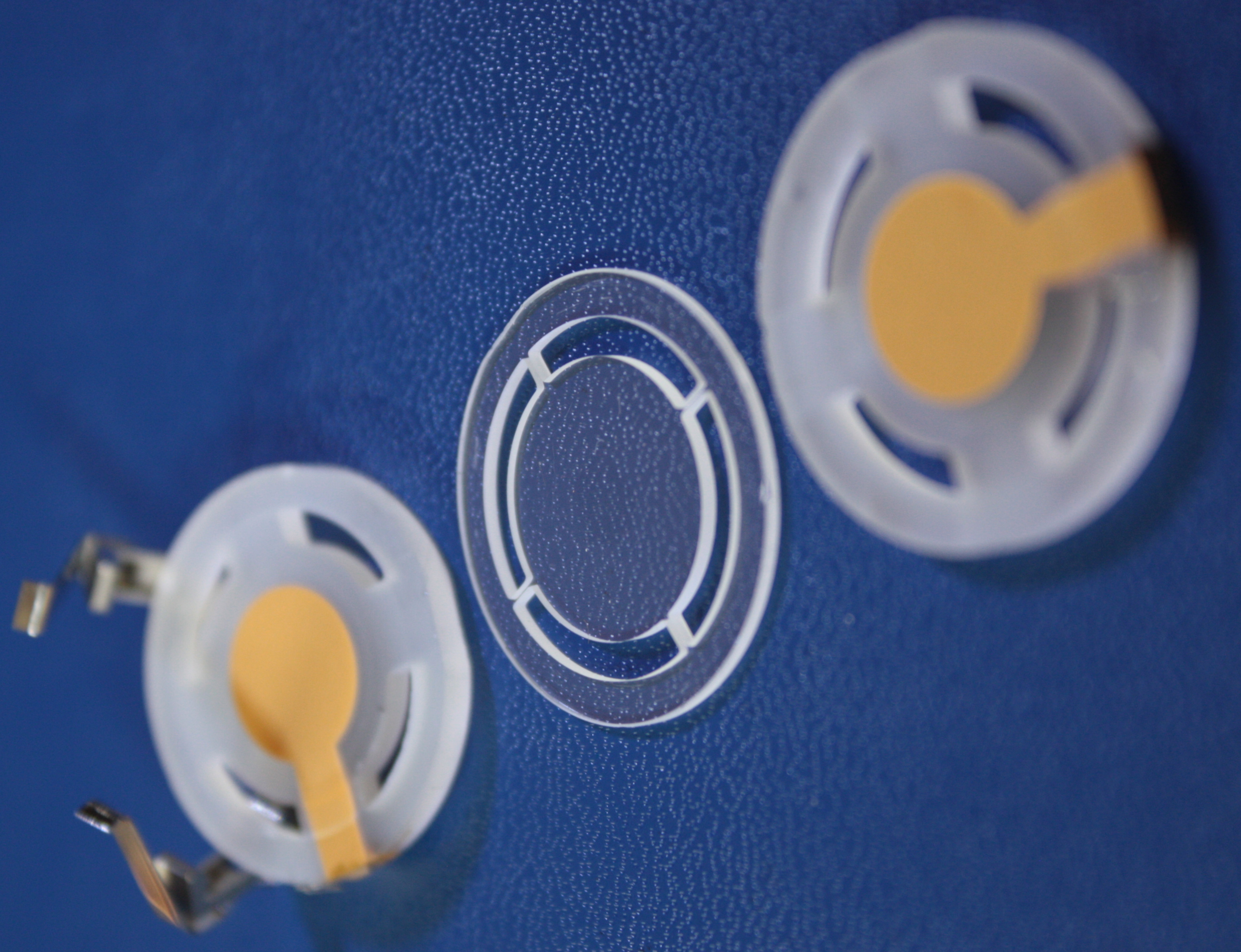}
	\caption{Bulk Acoustic Wave cavity: the active part is the central disk suspended to a rim by 4 "bridges". This plate is clamped between two quartz parts supporting the electrodes. Supports are a few micrometers far from the active part. }
	\label{fig:Device}
\end{figure}

The device under test, a BAW cavity, is an electrodeless version of a plano-convex quartz crystal resonator as shown in the center of Fig.~\ref{fig:Device}. It is made in a premium-quality quartz crystal slice in accordance with the so-called doubly rotated SC-cut (for "Stress Compensated", corresponding to rotation angles $\phi =22.4^\circ$, $\theta = 34.0^\circ$) exhibiting a low stress-to-frequency sensitivity on its metrologic mode, the slow thickness shear mode or C mode. The central disk is $1$~mm thick at its center, and its diameter is $15$~mm. The energy trapping is then optimized on the $3^\text{rd}$ overtone (OT), more precisely the (3, 0, 0) mode, of the C-mode at $4.9999$~MHz at room temperature (RT). Vibration frequencies of the $3^\text{rd}$ OTs of the fast thickness-shear mode, the B-mode, and the longitudinal thickness mode, the A-mode, are located at $5.47$~MHz and $9.31$~MHz respectively. It may be noticed that all these three  mechanical thickness modes are piezoelectrically coupled to an electrical field normal to the quartz plate in such a SC-cut whereas this is not the case for the well-known AT-cut for example. Typically all the odd OTs could be excited with electrodes deposited on a supported structure (Fig.~\ref{fig:Device}). Both electrode supports are also shaped in accordance with the plano-convex active disk to confine the vibration at its center. This dedicated device has the advantage of being quite easy to disassemble to coat one or both surfaces of the vibrating plate.
Although the A and B modes are extremely sensitive to temperature at RT (typically more than $-5 \times 10^{-5}$K$^{-1}$ at $300$~K for the $3^\text{rd}$ overtone (OT) of the B-mode), making them unusable in metrology applications, the $3^\text{rd}$ OT of the C-mode exhibits a rather weak temperature sensitivity, close to $+4\times 10^{-8}$ K$^{-1}$ at $300$~K, making frequency-shift measurements still achievable with a minimum of precautions even without a fine temperature control. But, in contrast, around $4$~K the fractional frequency sensitivity to temperature changes remain typically limited to a few $10^{-9}$K$^{-1}$ for all modes and OTs. So, a temperature control to within $10$~mK at these low temperatures makes relevant the comparison of frequency behaviors of all acoustic modes before and after coating of the quartz resonator.

\subsection{The graphene layer} 

The quartz resonator - just the part at the center of Fig.~\ref{fig:Device} - has been shipped to a high-quality graphene producer offering custom manufacturing services for graphene-based devices~\cite{Grapheneawebsite, Graphenea}. A $5$~mm diameter graphene disk has been transferred to the convex side of the resonator by the manufacturer himself, following his own chemical vapor deposition (CVD) standard transfer process in a class 1000 clean room. According to the provider, the graphene layer is first grown on a copper foil by the CVD method (in a cold walled CVD reactor at $1000~^\circ C$ and low pressure using methane as the carbon source, the copper foils being annealed at $1000~^\circ C$ under a hydrogen and argon flow before the graphene growth). A poly methyl methacrylate (PMMA) support layer is then spin-coated onto the graphene before chemical etching of the copper foil by a solution containing ferrite chloride. Finally, the resulting bi-component sheet is transferred onto the quartz substrate and the sacrificial PMMA layer removed by heating the sample at $450~^\circ C$ in inert atmosphere for $2~h$. Each batch is checked by means of a Raman spectroscopy and optical microscopy inspection to ensure a good transfer quality and purity, and the provider specifies that the monolayer is typically $0.35$~nm tick with a grain size up to $10\:\mu$m.  

\subsection{Method}
  
The characterization of graphene coating effects was made in two steps. Firstly, the device under test (DUT) was measured in its nominal configuration (no coating on the vibrating part) at $4$~K. Secondly, the BAW cavity was tested with the $5$~mm diameter graphene monolayer on a face. Additionally, as a reference test, the same BAW device has been used with gold and chromium coatings on both sides successively in order to check the process, and to compare their respective effects on the resonances~\cite{RSIcoating}. The device before and after graphene coating is characterized in terms of quality factors (inverse of mechanical losses) at resonance frequencies of overtones of the three thickness modes, according to a well-defined procedure~\cite{RSIcoating}.

\begin{figure}[h!]
	\centering
	\includegraphics[width=3.5in]{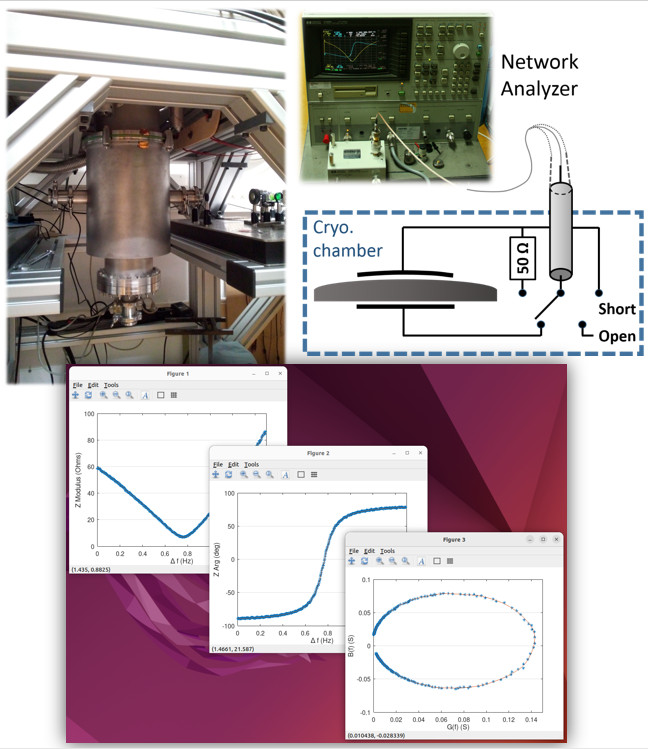}
	\caption{Parts of the experimental set-up and data extracted from. A HP4195A or Agilent 4395A-type network/spectrum analyzer with its impedance test kit is used to read the impedance or admittance of the DUT close to the resonance frequency of interest and record the data points from. This analyzer should be locked on a H-Maser to get a reliable frequency value. The DUT set on the cold stage of the pulse-tube cryocooler, a SHI RP-082B unit, is temperature controlled close to $4\:K$. The DUT, inside the cryo-chamber under vacuum, is at one end of a coaxial cable starting from a feed-through connector at room temperature at the other end, and three identical cables ended respectively by a $50\:\Omega$ load, a short-circuit and an open circuit are used to calibrate the system. Recorded data are typically the impedance (modulus and argument) and/or the admittance "circle" so-called GB plot (See \cite{RSIcoating} for details).}
	\label{fig:Exp_Setup}
\end{figure} 

In short, the method is based on measuring the bandwidth and/or equivalent electrical parameters of the device with a network analyzer locked on a Hydrogen Maser while the device is temperature-controlled around $4$~K in a commercial pulse-tube cryorefrigerator by means of a Lakeshore controller. The H-maser reference frequency exhibits a short-term fractional frequency stability of $1 \:10^{-13}$ over $1$~s combined with a long-term stability of $5\:10^{-16}$ over $10,000$~s. In addition, the laboratory is also connected to the French primary frequency standard (Observatoire de Paris - SYRTE) to guarantee the frequency accuracy. The analyzer span can be minimized down to $0.5$~Hz leading to a resolution close to $1.25$~mHz, and the sweep time as slow as $10.5$~mn is compatible with the expected unloaded Q values. A calibration should be done before measuring, as illustrated in Fig.~\ref{fig:Exp_Setup}, and the driving power is kept as low as possible in order to limit the power dissipated in the resonator to about $1$~nW. Q-factors are extracted from the recorded data as $Q=\frac{f_0}{f_H-f_L}$ or  $Q=\frac{f_0}{2}\frac{d\phi}{df}(f=f_0)$ where $f_0$ denotes the measured center frequency, $f_H$ the high cut-off frequency, $f_L$ the low cut-off frequency (i.e. $f_H-f_L$ is just the frequency bandwidth), and $\phi$ is the impedance argument. $f_0$, $f_H$, and $f_L$ can be easily measured from the impedance argument (Z Arg in Fig.~\ref{fig:Exp_Setup}) or conveniently from the "GB plot" - the imaginary part B of the admittance against its real part G - where $f_L$ and $f_H$ are respectively the frequencies at the maximum and minimum of $B(f)$ and $f_0$ is at the maximum of $G(f)$.

\section{Results and Discussions}
\subsection{Mechanical loss of a supported graphene-layer at $4$~K}
\label{sec:Qfactor}
\begin{figure}[h!]
	\centering
	\includegraphics[width=3.5in]{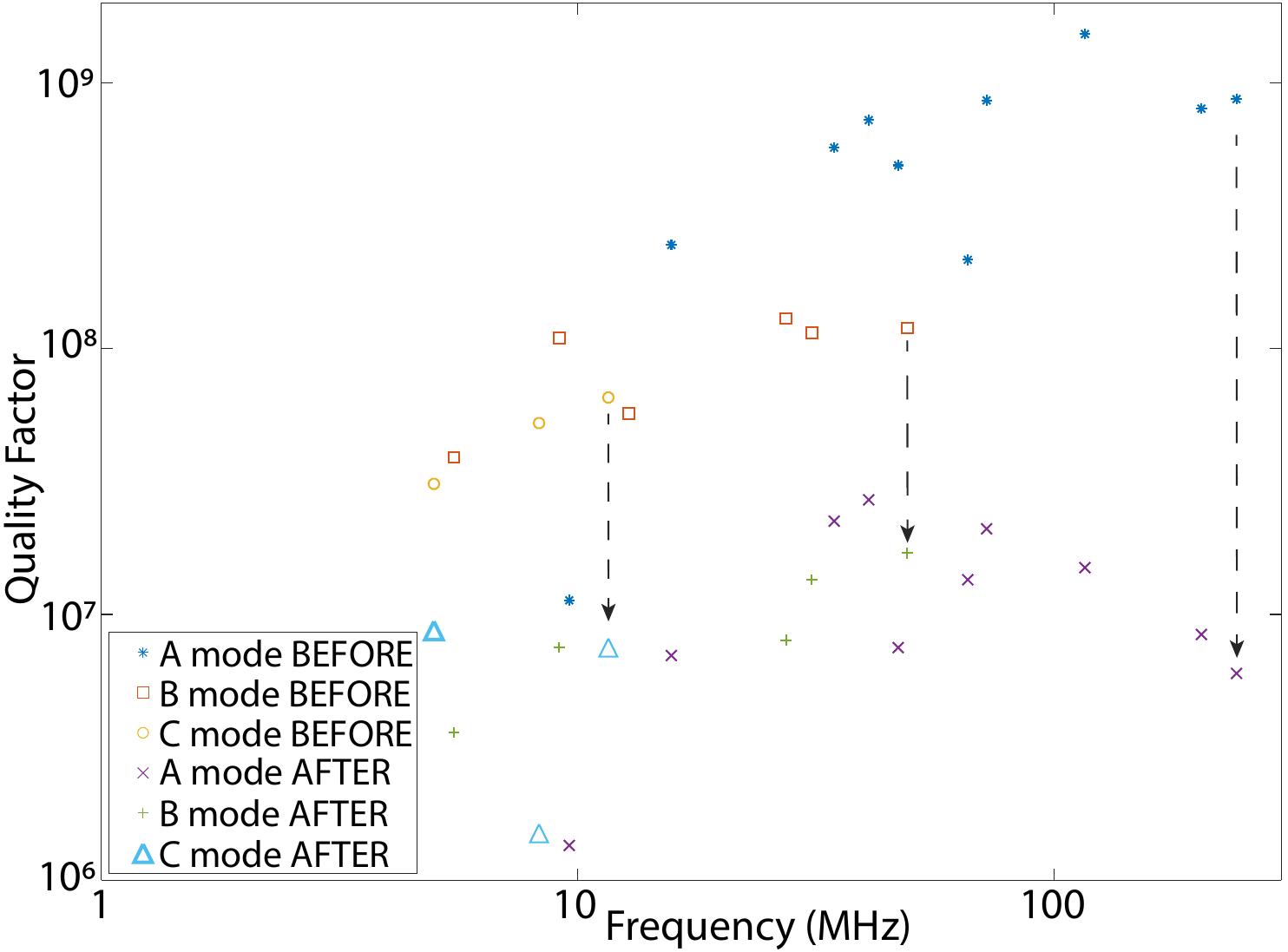}
	\caption{Quality factor versus frequency for different OTs of the three vibration modes before and after graphene monolayer coating measured at $4$K.}
	\label{fig:Q_vs_freq}
\end{figure}

Low loss acoustic cavities can be used to probe mechanical losses in various coatings\cite{RSIcoating}. Indeed, total losses of a coated device is, ideally, a sum of intrinsic losses of the acoustic plate and the coating material. So, by comparing quality factors of these devices before and after coating, one can deduce material properties of the added layer. Thus, since the BAW resonator internal losses set limits on the detectable effects, it is straightforward to discuss the main dissipation mechanisms limiting BAW performance. For frequencies typically greater than a few Megahertz and at room temperatures, BAW devices operate in the Akhieser regime~\cite{Akheiser} which corresponds to the well-known $Q\times f = \text{const.}$ dependence between losses and wave frequency $f$. On the other hand, for temperatures $T$ close to $4$~K, same devices operate in the Landau-Rumer regime~\cite{landaurumer1}, because the thermal phonon lifetime is $1/\tau_{th} < f < k_{B}T/\hbar$. In this regime the acoustic wave absorption coefficient $\alpha(f)$ is proportional to $T^n f$ with $n$ close to 4 or 6 depending on whether the acoustic wave is a shear one or longitudinal~\cite{landaurumer1, Maris1971}. Consequently, the $Q$-factor becomes independent of the frequency~\cite{Gory13} because $Q \propto \frac{f}{\alpha(f) V}$ where $V$ is the wave velocity. Although these relationships are true for intrinsic losses linked to a three phonon mechanism, in practice, additional engineering losses may lead to deviations from this law. As shown in Fig.~\ref{fig:Q_vs_freq}, experimental data exhibit two trends, even for the bare resonator (plots labeled "before", for "before coating"): at the lowest frequencies Q-factors remain limited by energy trapping imperfections whereas surface scattering occurs at higher frequencies, here from about $115$~MHz, because of the residual roughness of the polished surfaces (a few nm typically). In any case Q-factors drop down once the resonator is graphene-coated (see plots so-called "after" for "after coating" in Fig.~\ref{fig:Q_vs_freq})
%

Ideally, additional loss of a deposited layer, e.g. graphene, can be estimated from the Young moduli~\cite{Berry1981} of both the substrate, in this case crystalline $\alpha$-quartz, and the coating~\cite{RSIcoating}, assuming that intrinsic losses are dominant and that the interface damping is negligible. Indeed, neglecting the weak anisotropy and piezoelectric of quartz, resulting losses in the coated device can be simplified as: 
\begin{equation}
\label{eq:loss_equation}
\Phi_{\text{coated}-q} \approx \Phi_{q}+\frac{\mathcal{E}_{g}}{\mathcal{E}_{q}}\Phi_{g}\approx \frac{\mathcal{E}_{g}}{\mathcal{E}_{q}}\Phi_{g} \approx \frac{3 t_{g}Y_{g}}{t_{q}Y_{q}}\Phi_{g},
\end{equation} 
where $\Phi$ denotes mechanical loss ($\sim 1/Q$), $\mathcal{E}_g$ ($\mathcal{E}_q$) is the energy stored in graphene (quartz), $Y_i$ Young's moduli, and $t_i$ the thicknesses. A graphene Young modulus along the layer plane of $1$~TPa has often been reported at room temperature~\cite{Bunch}. That of quartz is estimated to be $86$~GPa~\cite{RSIcoating}.
 
From the experimental data plotted in Fig.~\ref{fig:Q_vs_freq}, by extrapolating the calculation to the best case achieved with the A-mode at $115$~MHz, the frequency from which surface scattering occurs, the $1$~mm thick quartz coated with a $0.35$~nm thick layer of graphene would exhibit a mechanical loss $Q_{\text{coated}-q}^{-1}=\Phi_{\text{coated}-q} \approx 9\times 10^{-7}$. Thus, the graphene layer loss at $4$~K would be estimated from Eq.~(\ref{eq:loss_equation}) as close to $\Phi_g \approx 8\times  10^{-4}$, in the best case, with an uncertainty mainly linked to that of Young modulus and thickness of a graphene monolayer at $4$~K. Similar values have been observed for gold and chromium coatings under the same operating conditions and with the same device under test: $\Phi_{Au}\approx 4\times 10^{-4}$, and $\Phi_{Cr}\approx 16\times 10^{-4}$ respectively with the latter depending on frequency~\cite{RSIcoating}.

The above graphene-loss assessment is greater than those reported for micro-scale suspended monolayers, doubly-clamped~\cite{Chen2009, Takamura2016} or clamped-on-all-side suspended membranes~\cite{VDZ2010, Takamura2016}, typically $1-1.4\; 10^{-4}$, but this could just be attributed to the larger surface of adhesion involved in the present case.
Nevertheless, losses of $Q_g^{-1}=\Phi_g \approx 8\times  10^{-4}$ are also much greater than those measured at $4$~K with another film-on-substrate device, a single-layer graphene film deposited on the so-called "double paddle oscillator (DPO)"~\cite{LiuNanoLett}, for which internal friction $Q_g^{-1}$ of less than $0.3\:10^{-4}$ are mentioned. Even with thicker multilayers on such a DPO, graphene still exhibits losses as low as $3.1\:10^{-4}$ and $2.6\:10^{-4}$ for CVD graphene coatings of respective thicknesses $8$~nm and $6$~nm~\cite{LiuDiamond, Takamura2016}. 
The tested DPO is also a $mm$-scale system coated with an exfoliated CVD graphene film like for our DUT but differs from it by the substrate nature, $SiO_2/Si$ instead of $\alpha-$quartz crystal, and the  operating vibration, torsion at low frequency (typ. $5.5$~kHz) instead of $MHz$ shear or expansion modes in our resonator.

How can such a discrepancy of a graphene-loss value be explained? Beyond the dispersion of mechanical coefficients - often larger for the shear modulus of a single-layer graphene than for its Young modulus $Y_g$, for example~\cite{Liu2Dmat2021, LiuNanoLett} - actually, intrinsic losses also depend on the stress fields in both materials, graphene and substrate. When operated at $4$~K, thermomechanical stresses appear inevitably in such heterostructures assembled at RT, and obviously differ from a BAW quartz resonator to a $SiO_2/Si$ DPO in torsion. This point about existing stresses in the graphene-coated quartz-resonator at $4$~K is discussed below in a dedicated paragraph.
Extra losses could also come from the graphene-substrate interface involving Van Der Waals forces typically, and again would depend on the nature of the substrate~\cite{LiuNature2019, WeiJAP2020, QiuAPL2012} (and/or to a possible annealing process).

In addition, Q-factors could also be degraded by an engineering loss originating from an imperfect centering of the deposited graphene film. Indeed, the graphene "sticker", the circular graphene film, is transferred manually onto the plano-convex quartz disk, making this operation critical among possible manufacturing defects. Such a defect similar to a off-center mass loading could couple a unperturbed mode of interest $(n,0,0)$ with a odd-symmetry anharmonic mode $(n,p,0)$, typically a $(n,1,0)$ mode, $p$ being odd, assuming that the off-center mass perturbation is after $x_1$~\cite{EernisseTUFFC1990}.

In Fig.~\ref{fig:Q_vs_freq} we can also observed a shift of the corner frequency marking the Q-factor decrease due to a degradation of the surface roughness, leading to wave scattering when the frequency increases and therefore an increase in losses: this frequency, close to $115$~MHz for the A mode when there is no graphene (the surface roughness being about $4$~nm), changes to about $41$~MHz ($49$~MHz for the B-mode) once the graphene layer is in place. Therefore,  with a graphene-monolayer coating at $4$~K the roughness standard-deviation~\cite{galliouScRip2013} of the resonator becomes $\sigma=\frac{t_q}{\sqrt{2nQ}} \approx 35\:nm$, $n$ being the OT order and $Q$ the corresponding Q-factor, i.e. the 13th OT of the A mode exhibiting a Q-factor of $27\:10^6$ at $40.8$~MHz (the 27th OT of the B-mode with $Q \approx 17\:10^6$ at $49$~MHz).

\begin{figure}[h!]
	\centering
	\includegraphics[width=3.5in]{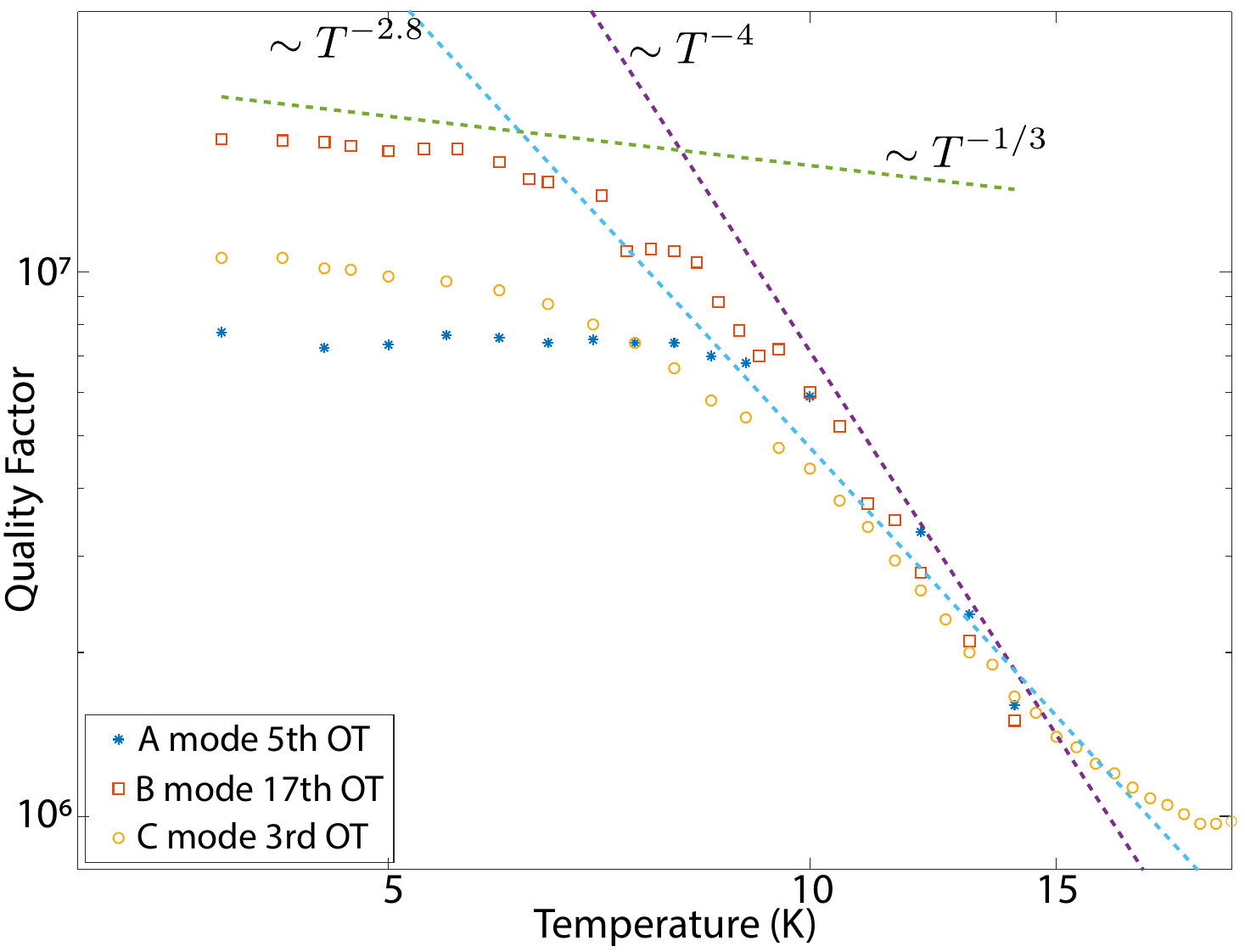}
	\caption{Q-factor versus temperature for a selection of OTs of the three vibration modes after graphene monolayer coating.}
	\label{fig:Q_vs_Temp}
\end{figure}

Regarding Q-factor behaviors with temperature, trends shown in Fig.~\ref{fig:Q_vs_Temp} suggest that losses for $T>4$~K are limited by phonon-phonon interactions corresponding to the Landau-Rumer regime, because $Q$-factors scale as $T^{-n}$. Nevertheless, the exponent $n$ is less than 4 instead of typically $4\leq n \leq 6$~\cite{Lewis, Maris1971}. For lower temperatures, a $T^{-1/3}$ scaling law could be attributed to residual impurities in the synthetic quartz crystal generating TLS~\cite{Gory12,galliouScRip2013}, but is not systematic depending on the mode considered.  
\subsection{Frequency shifts due to graphene mass-loading}
\label{sec:massloading}
In addition to the $Q$-factor measurements, effects of a coating can also be characterized by the corresponding frequency shifts appearing as another possible source of information. 

The simplest mechanism that may cause a frequency shift is the mass loading (ML) effect: adding an extra layer of material increases the effective mass of acoustic modes leading to decrease in frequency which is inversely proportional to the mass. This effect is commonly used to tune the resonance frequency of electroded devices or to detect an extra mass (see QCMs). It is important to note that the graphene layer cannot resonate by itself because its thickness is much lower than half of the acoustic wavelengths concerned in this work. Due to this effect, the frequency shift of an acoustic mode resonating at $f_{n0}^{'}$ can be estimated as:
\begin{equation}
\frac{\Delta f_n^{'}}{f_{n0}^{'}} \approx -\frac{\rho_l t_l}{\rho_q t_q},
\label{eq:fractionalfreq with graphene mass loading}
\end{equation} 
where $\rho_q$ ($\rho_l$) is the mass density of quartz (layer),  $t_q$ ($t_l$) is the thickness of quartz (layer). This shift resulting of approximations does not depend anymore on the vibrating mode type, A, B or C, at the first order (See details in \ref{sec:appendixmassloading}. This approximation known as Sauerbrey formula is very popular in the QCM community~\cite{Sauerbrey, Lostis, Lu, Mecea, Reed, Johannsmann,JohannsmannBook}). The areal mass of a graphene coating can be assessed as $\rho_l t_ l = N m \approx 7.6\:10^{-4}\: g/m^2$, where $N$ is the number of atoms per unit of area and $m$ the atom mass (when considering 2 full carbon atoms per C-hexagon whose C-C length is 0.142 nm), and would induce, ideally, a frequency shift of $-2.85\:10^{-7}$.

At $4$~K, the fractional frequency shift between the uncoated and graphene-coated resonator can be calculated as a function of the frequency shifts at $300$~K, here denoted $-R_{300K}$ for the expression in Eq. \ref{eq:fractionalfreq with graphene mass loading} at $300$~K, and integrated coefficients of thermal expansion (ICTE) from $300$~K down to $4$~K as:
\begin{eqnarray}
\frac{\Delta f_{ML_{4K}}}{f_{4K}} &\simeq&-\frac{\rho_g t_g}{\rho_q t_q}=-R_{300K}\frac{(1+\alpha_i\delta T)(1+\alpha_g\delta T)}{(1+3\alpha_g\delta T)(1+\alpha_2\delta T)} \nonumber\\
&\approx&-R_{300K}[1+(\alpha_1-\alpha_g)\delta T+(\alpha_3-\alpha_g)\delta T]
\label{eq:grapheneMassLoadingat4K}
\end{eqnarray}
where $\alpha_j= \alpha_j(T)$ denotes coefficients of thermal expansion (CTE) at a temperature $T$. Comparing results at $4$~K and $300$~K, infinitesimal component $\alpha_j(T) \delta T$ should be replaced with the corresponding integrated version (ICTE) over the temperature range: $\int_{T_0}^{T} \alpha_i(T) \mathrm{d}T $, $T_0=300$~K. Ref~\cite{Barron} provides relevant values for the integration of quartz expansion coefficients, giving $\alpha_1\delta T=\alpha_2\delta T=-2.54\times 10^{-3}$, $\alpha_3\delta T=-1.24\times 10^{-3}$ for quartz crystal within the considered temperature range. Estimations of ICTE for the graphene layer varies depending on the reference source: it is $\alpha_g\delta T =+1.1\times 10^{-3}$ from data by Ref~\cite{Mounet}, ~\cite{Vibhor} whereas it is closer to $+3.7\times 10^{-3}$ from data by Ref~\cite{Yoon}. It should be noted that graphene expands when cooled down while quartz contracts. As a result of Eq.~(\ref{eq:grapheneMassLoadingat4K}), the fractional frequency shift at $4$~K for a graphene-coated quartz would be again close to $-3\times  10^{-7}$ (i.e. $-R_{300K}$ multiplied by $+0.9936$ or $+0.9885$ depending on the ref. source). 

\begin{figure}[h!]
	\centering
	\includegraphics[width=3.5in]{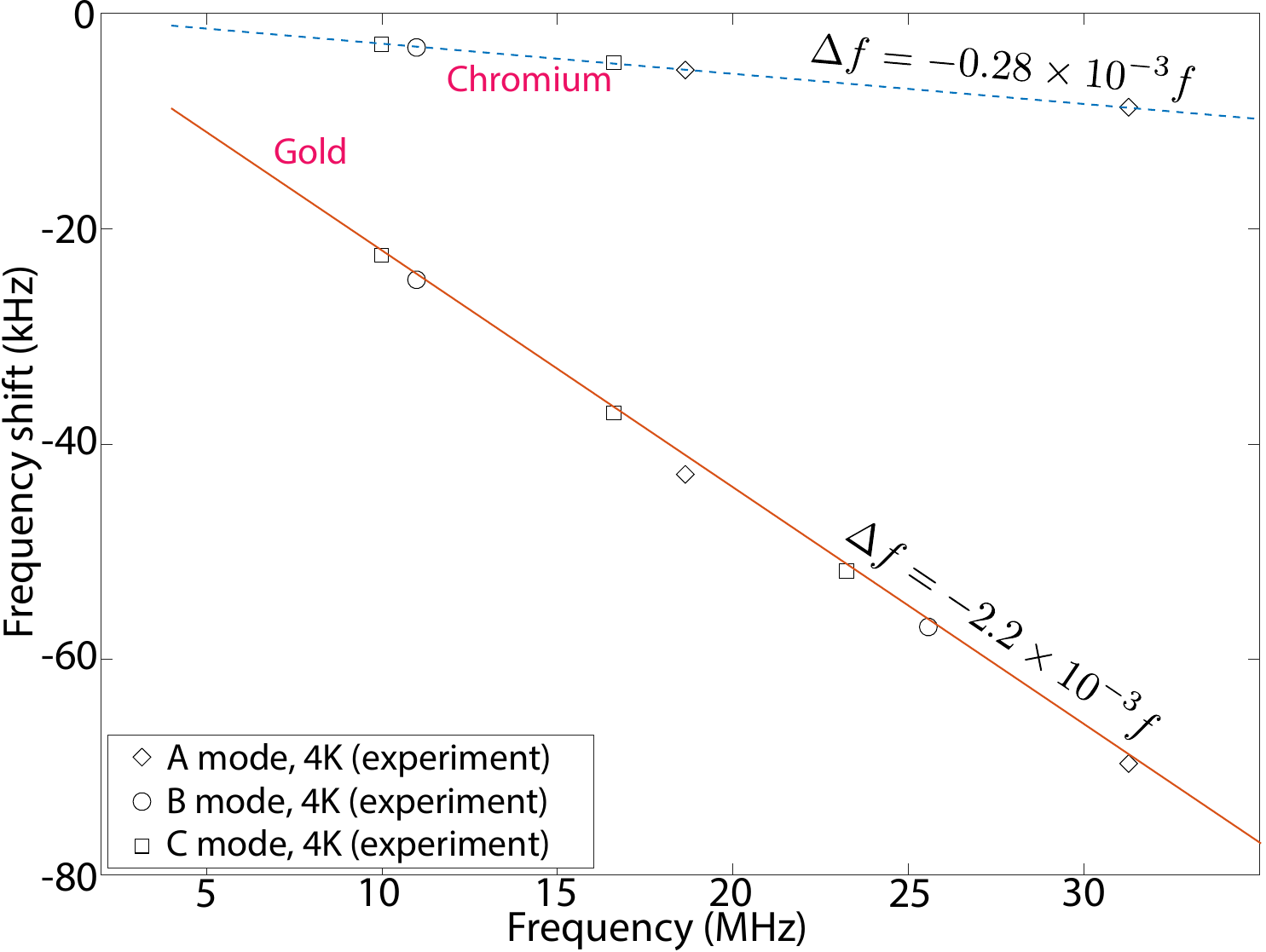}
	\caption{Frequency shift (difference between coated and uncoated cases) as a function of the frequency for various OTs of the three vibration modes for gold and chromium coatings on both sides of a plate measured at $4$~K. Coatings are $50$~nm thick for Cr and $150$~nm for Au. Solid and dashed lines correspond to the theoretical mass-loading effect.}
	\label{fig:Au_Cr_mass_loading}
\end{figure}

To check the methodology described above, additional tests have been carried out previously at $4$~K with more traditional gold and chromium coatings (whose mechanical and thermal properties are better known that those of graphene from RT to $4$~K):  first a $50$~nm thick chromium coating, and second a $150$~nm-thick gold over a similar area of $6$~mm diameter (both thicknesses are typically used in electroded quartz crystal resonators)~\cite{RSIcoating}. In both cases coatings were used as excitation electrodes. It should be noted though that films with such thicknesses exhibit properties, especially CTE, not so far from those of bulk materials~\cite{Oh}. Thus, since thin film properties at $4$~K are not known, gold and chromium ICTE can be estimated from bulk material data~\cite{White, NBS} as $-3.3\times 10^{-3}$ and $-9.8\times 10^{-4}$ respectively. The corresponding estimates of fractional frequency shifts $\frac{\Delta f_{ML_{4K}}}{f_{4K}}$ are then about $-2.2\times 10^{-3}$ and $-0.27\times 10^{-3}$ respectively.
These theoretical assessments of mass loading effects for Au/Cr coatings are compared with their corresponding experimental results in Fig.~\ref{fig:Au_Cr_mass_loading}, showing that the latter can well be fitted with linear functions of frequency $f$ in good agreement with the calculated values from Eq.\ref{eq:grapheneMassLoadingat4K}. It is therefore demonstrated that the mass-loading effect dominates in these cases and that it does not depend on the vibration mode.

\begin{figure}[h!]
 	\centering
 	\includegraphics[width=3.5in]{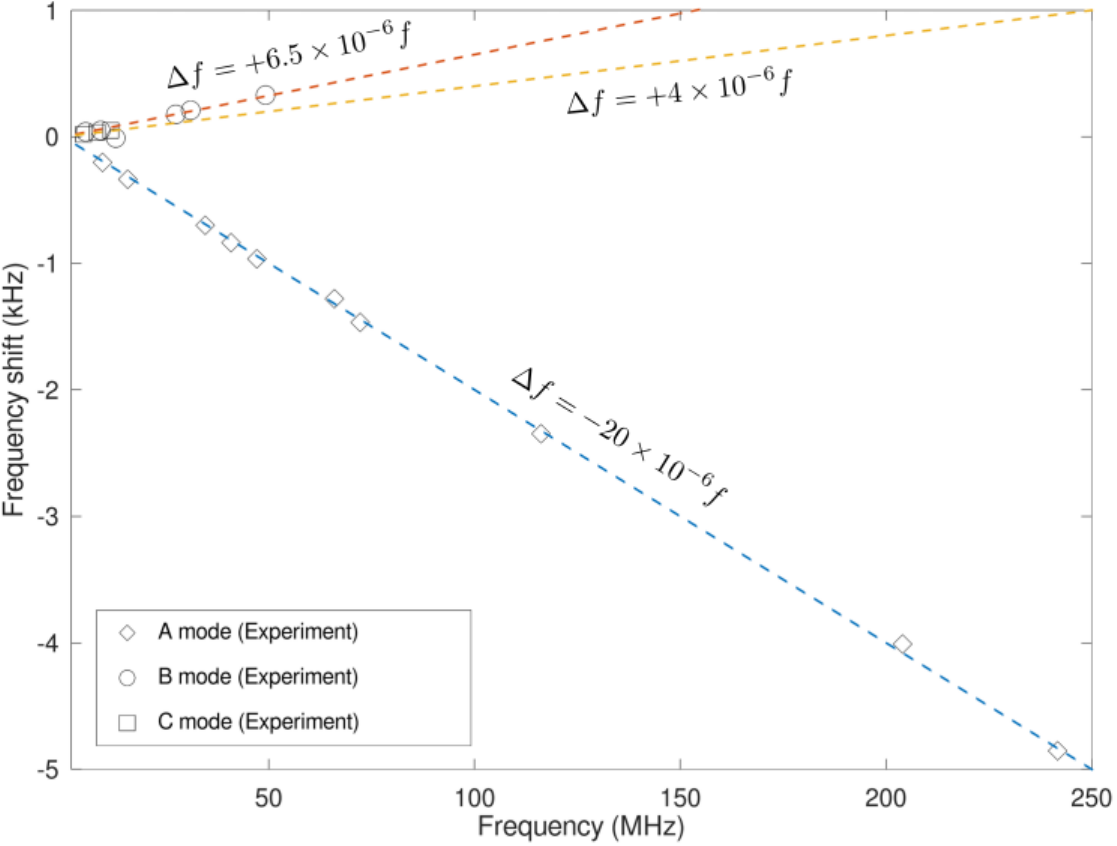}
 	\caption{Measured frequency shift $\Delta f$, difference between coated and uncoated cases, as a function of frequency $f$ at $4$~K, for various OTs of the three vibration modes for a graphene monolayer coating whose theoretical thickness is $0.35$~nm. High-order OTs of the shear modes cannot be measured due to relatively low $Q$-factors. Dashed lines are best fits.}
 	\label{fig:graphene_mass_loading}
 \end{figure}

 However, as shown in Fig.~\ref{fig:graphene_mass_loading} in the case of a graphene coating, behaviors are rather disappointing by taking into account only this effect of mass loading. Indeed, both shear modes exhibit a positive frequency shift proportional to the overtone number $n$, and the negative slope of the longitudinal mode significantly deviates from the expected mass ratio of graphene coating and quartz. As a result, although the methodology works, the mass loading model does not hold anymore in the case of a graphene coating.
  
To extend the modeling, one might add viscoelasticity of the coated film. This involves the ratio of Young modulus weighted by their respective densities~\cite{Johannsmann, JohannsmannJAP}.  Although the corresponding correction term remains negligible. Additionally, some other typical QCM modifications in a small load approximation have also been considered, keeping in mind that adhesion of graphene is strong~\cite{Lee, Koenig, Deng}. Among them friction modeled by a spring without any inertial effect or a spring with a dash pot to take into account losses. These modifications could explain a positive slope of frequency shifts versus the overtone order. Such positive frequency shifts of "composite resonator" have already been reported in rather specific cases~\cite{Castro, GalliMarxer, Pomorska}, although they do not match very well to the case of a graphene layer. Indeed, such a spring-type coupling gives $ \Delta f \propto +\frac{k_s}{m_q f_0}\frac{1}{n} $, where $k_s$ is a spring constant. Although the slope sign is positive, it has a $n^{-1}$ scaling which is difficult to verify experimentally because of the $Q$-factor decrease with the overtone order $n$ for both the C and B modes.
\subsection{Stress induced frequency shifts}
A more realistic additional effect that could, at least partially, improve predictions of the model is that of static thermomechanical stresses which definitely exist in this composite device due to a mismatch of graphene and quartz thermal expansion coefficients. 
Indeed, tests are performed at cryogenic temperatures while the graphene monolayer is deposited on the quartz substrate at room temperature (RT) according to a nominally stress-free(?) process.
Graphene exhibits a negative thermal expansion coefficient~\cite{Yoon} whereas that of quartz along the x-axis is always positive~\cite{Barron}, and that along the z-axis becomes negative between $5$ and $12$~K.
Due to this mismatch, the quartz plate bends because the graphene film is coated only on one side. In a SC-cut, the associated stress gives frequency shifts that are consistent with our experiment data (Fig. \ref{fig:graphene_mass_loading}) and supported by other arguments. 
Firstly, works by Ballato, Eernisse, and others show that stress induced frequency shifts are proportional to the operating frequency. 
Secondly, theoretically the A-mode shifts happen in opposite sign when compared with C and B mode deviations with respect to azimuth angle~\cite{BallatoUS1978}. 
Thirdly, C-mode frequency shift observed experimentally is much lower in absolute values than that for the A mode since the SC-cut plate is optimized to exhibit low stress sensitivity of the C-mode at RT.

Effects of a static mechanical bias on elastic waves, i.e. small dynamic fields superimposed on a static bias, were intensively studied in 70's-80's~\cite{Tiersten1978, Baumhauer1973, Sinha1982, Tiersten1981} after Thurston and Brugger works in 1964 ~\cite{ThurstonPhysRev1964}. In this work, we employ Sinha-Tiersten's perturbation analysis limited to the perturbation of the elastic constants and not including dielectric or piezoelectric constant changes for example which can be justified by the weak piezoelectric coupling of quartz~\cite{Stevens1997}.

\begin{figure}[h!]
	\centering
	\includegraphics[width=3.25in]{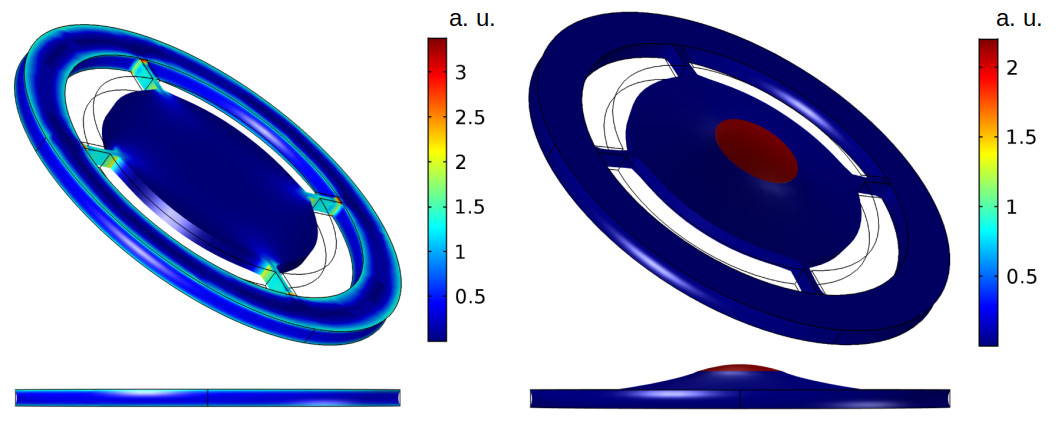}
	\caption{Von Mises stresses calculated by FEM simulations for the resonator clamped at its rim. To the left: the uncoated quartz resonator cooled down from $300$~K to $4$~K, with its side view at the bottom. To the right: the graphene-coated resonator with its side view at the bottom}
	\label{fig:FEMresults}
\end{figure}

Details on our calculation process are given in \ref{sec:appendixstaticStresses}. Numerical values have been taken from Ref~\cite{Bechmanncoef21958} for piezoelectric and stiffness coefficients at RT (See also Table \ref{tab:Dataat300K} for useful data at RT), and from Ref~\cite{Tarumi2007} for the same coefficients at $4$~K.  In a preliminary step, quartz ICTEs, $\alpha_i\delta T$ integrated over $[300\:K, 4\:K]$, have also been checked based on values from Ref~\cite{Barron} calculated for our doubly-rotated quartz cut: the effective elastic constants $\bar{c}_{4K}$ had to be adjusted by less than $2\%$ so that the calculated A-B-C-mode fractional frequency shifts meet the experimental values.
Otherwise, the calculation process is based on the relationship between the frequency shift resulting of static stresses, or their related strains, in the vibrating thickness through a perturbation tensor. The latter can be calculated from a reference state at $4$~K by means of the set of parameter values from Ref~\cite{Tarumi2007} applied to a doubly rotated quartz cut. The calculation process can be summarized as follows:


\begin{enumerate}[label=\alph*)] 
	\item The uncoated resonator can be seen as a circular plate with radius $r_q$ and tickness $t_q$ subjected to an extra diametrically applied force $F$ in the plane $(x_1, x_3)$ coming from constrained contractions of its four bridges induced by cooling from RT to $4$~K. Assuming that the resonator rim is clamped, the naked device would exhibit a fractional frequency change (See \ref{sec:appendixstaticStresses}):
	\begin{eqnarray} 
	\frac{\Delta f_{4K}}{f_{4K}} \simeq \frac{1}{t_q \rho_q v^2} \frac{\sigma_i t_q K_i}{2} 
	\simeq \frac{\sigma_i}{2} R_i
	\label{eq:dfoverfcalculated2}
	\end{eqnarray}
	for $i = 1, 3 ,5$, and with:
	\begin{equation}
	K_{me} = 2c_{2\alpha 2n}s_{n\gamma me}V_{\alpha}V_{\gamma} +c_{2\alpha 2\gamma ab}s_{abme}V_{\alpha}V_{\gamma} + \delta_{2m}\delta_{2e},
	\label{eq:Ki}
	\end{equation}
	$c_{2\alpha 2n}$ and $s_{n\gamma me}$ being elastic coefficients, $V_{\alpha}$ eigenvectors ($R_i=\frac{K_i}{\rho v^2}$ are sometimes known as Ratajski coefficients~\cite{Ratajski}). Obviously eigenvalues $\rho_q v^2$ and constants $K_i$ depend on the mode of interest, A, B or C, and all coefficients are calculated for the doubly-rotated SC-cut at $4$K: see Table \ref{tab:DataSCcut5K}. For a four-point mounting in the $(x_1, x_3)$ plane, stresses $\sigma_i$ at the center of a circular plate can be adapted from Ref.~\cite{Janiaud1978} to give $\sigma_5\simeq0$ for a SC-cut, while $\sigma_1\simeq \sigma_3 \simeq \frac{-2F}{\pi t_q r_q}$, $F$ depending on the ICTEs (see \ref{sec:appendixstaticStresses}).
	\item The resonator one-sided coated with a graphene layer is sensitive to the thermal expansion mismatch and to the diametrical force of its bridge-holders. Actually, the latter is very close to that of the uncoated resonator as shown in \ref{sec:appendixstaticStresses}. Regarding stresses induced by the thermal expansion mismatch of both materials, they can be simplified as linear functions of the thickness coordinate $x_2$~\cite{Janiaud1978} (FEM simulations as illustrated in Fig.~\ref{fig:FEMresults} confirmed this simplification) written as $\sigma_i(0, x_2, 0)=a_ix_2+b_i$, leading to a fractional frequency change, for $i = 1, 3 ,5$:
	\begin{eqnarray}
	\frac{\Delta f_{4K\_g}}{f_{4K}} \simeq \frac{1}{t_q \rho_q v^2} \frac{b_i t_q K_i}{2} 
	\simeq \frac{b_i}{2} R_i, 
	\label{eq:dfoverfcalculated2graphene_T}
	\end{eqnarray}
	where it is shown (\ref{sec:appendixstaticStresses}) that $b_1 = b_3 \simeq \frac{Y_g }{1-\nu_g}\frac{t_g}{t_q}(\alpha_g-\alpha_q)\delta T,$ $b_i \simeq 0$ otherwise, and where infinitesimal ${\alpha_j}\delta T$ should be replaced with corresponding ICTE: $\int_{300K}^{4K}\alpha_j \, \mathrm{d}T$.
	\item Then, because of the very similar effect of diametrical forces $F$ exerted by the quartz bridges in both previous cases, the resulting fractional frequency shift can finally be expressed as:
	\begin{equation}
	\frac{f_{4K\_g}-f_{4K}}{f_{4K}}\simeq \frac{R_i}{2}\frac{Y_g}{1-\nu_g}\frac{t_g}{t_q}(\alpha_g-\alpha_q)\delta T. 
	\label{eq:dfoverf_after_before_T}
	\end{equation}
	
\end{enumerate}

The above stresses involved in the fractional frequency changes lead to calculated values ranging from about 1 kPa to 10 kPa depending on the data used: $Y_g$ is often set to $1$~TPa but may be lower, reported Poisson coefficient $\nu_g$ are from $0.17$ to $0.78$, and the ICTE difference is from $2\:10^{-3}$ to $6\:10^{-3}$, the graphene thermal expansion coefficient $\alpha_g(t)$ being still discussed. It may be noticed that by including the (weak) anisotropy of quartz in the modeling (See \ref{sec:appendixstaticStresses}), stresses at the substrate center are, in comparison with the isotropic approximation, $b_{1aniso}\approx$ 98\% $b_{1iso}$, $b_{3aniso}\approx$ 96\% $b_{3iso}$, and $b_{5aniso}<-1.4\:10^{-3}$ instead of zero. Finite-Element-Method simulations (See Fig.~\ref{fig:FEMresults}) have been performed in parallel to check the analytical results. Stresses induced by the composite-device cooled down to $4$~K are simulated by using the set of CTE from Ref~\cite{Barron} for the quartz substrate and Ref~\cite{Yoon} for the graphene layer. These simulations provide numerical results similar to those obtained by the analytical modeling.
 
Nevertheless, the fractional frequency differences taken from Eq. (\ref{eq:dfoverf_after_before_T}) do not meet the measured ones (Fig.~\ref{fig:graphene_mass_loading}). Actually,
frequency shifts given by Eq. (\ref{eq:dfoverf_after_before_T}) have to be balanced by extra offsets including a mass-loading effect of a few $-10^{-7}$ to explain the experimental values in Fig.~\ref{fig:graphene_mass_loading}.

\begin{figure}[h!]
	\centering
	\includegraphics[width=3.25in]{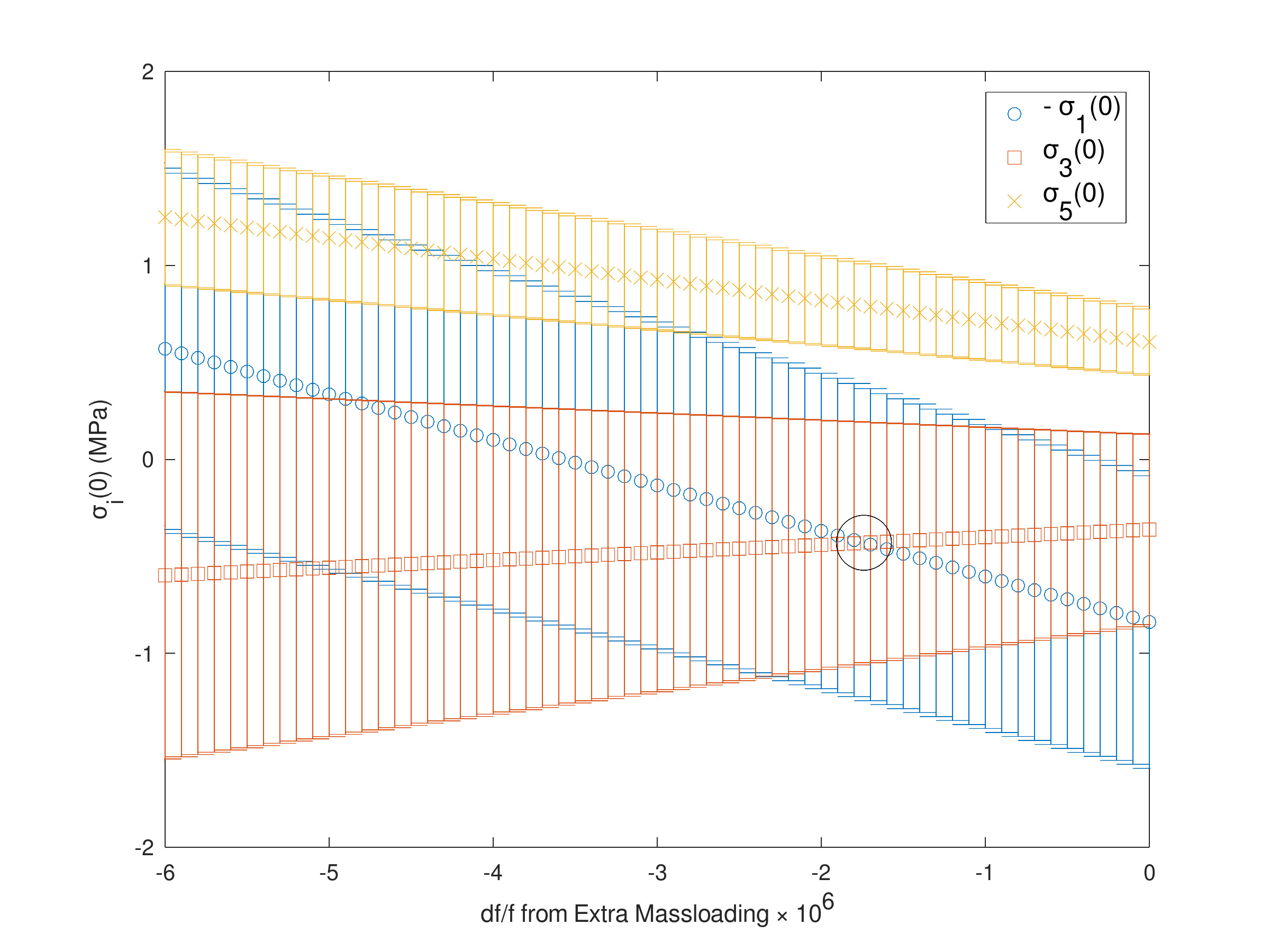}
	\caption{Calculated values of the thermo-mechanical stresses $\sigma_i(0)$ , $i=1, 3, 5$, at the center of the quartz substrate coated with an ideal graphene monolayer, as solutions $b_i=\sigma_i(0)$ of the system of Eqs.\ref{Eq:systeqn} as a function of a mass loading effect $\frac{\Delta f_{ML_{4K}}}{f_{4K}}$. Such a set of stresses, induced by a  CTE mismatch (but not only?) and combined with the mass loading effect, satisfies experimental frequency shifts recorded at $4$~K (Fig.\ref{fig:graphene_mass_loading}). Error bars result from a $5\%$ uncertainty applied to the force-frequency coefficients $R_{ij}$. The circle marks the most probable solution.}
	\label{fig:Stresses}
\end{figure}

Previous relationships can be used to converge toward a set of realistic values of mass loading on the one hand and induced mechanical stresses at the center of the composite device, $\sigma_i(0)=b_i$, $i=1, 3 , 5$, on the other hand, compatible with the three measured frequency shifts from Fig.~\ref{fig:graphene_mass_loading}. 
To do so, the issue consists in solving the set of three equations, one per vibration mode, with three unknown stresses $b_1, b_3, b_5$ ($b_i=\sigma_i(0)$), as a function of an unknown additional shift caused by mass-loading $\frac{\Delta f_{ML_{4K}}}{f_{4K}}$:  
\begin{equation}
\frac{f_{4K\_gx}-f_{4Kx}}{f_{4Kx}}=\frac{R_{1x}}{2}b_1+\frac{R_{3x}}{2}b_3+\frac{R_{5x}}{2}b_5+\frac{\Delta f_{ML_{4K}}}{f_{4K}}, \; \textrm{ with \textit{x}\,=\,A,\,B,\,C},
\label{Eq:systeqn}
\end{equation}
$R_{ij}$ being the corresponding force-frequency coefficients of each mode, and in the left hand side are put the respective experiment values from Fig.~\ref{fig:graphene_mass_loading}. Solutions $b_i=\sigma_i(0)$ are shown in Fig.\ref{fig:Stresses} within the range $- 60\times 10^{-7} \leq \frac{\Delta f_{ML_{4K}}}{f_{4K}} \leq 0$ corresponding to an added areal mass that could reach up to $15\: ng/mm^2$. The theoretical areal mass of a graphene monolayer being around $1\: ng/mm^2$, the extra mass involved here could just be due to pollution and/or contamination that could occur during the DUT installation into the cryorefrigerator, done in a laboratory environment and not in a clean room. The amount of dust on the graphene surface, once the device out of the cryogenic vacuum chamber, is estimated in \ref{sec:appendiximpurities}. The resulting order of magnitude is consistent with the areal mass mentioned above, although it is impossible to say that the amount of impurities present under vacuum, at 4K, is the same as that measured after the device is removed from the vacuum chamber! 

Quartz is not a very anisotropic material and therefore stresses should be such that $|\sigma_1(0)| \approx |\sigma_3(0)|$ and $\sigma_5 \approx 0$ in the ideal case simulated here. As shown in Fig.\ref{fig:Stresses}, these conditions are far from being met: especially $b_5=\sigma_5(0)$ reaches unexpected orders of magnitude and values of $\sigma_1(0)$ $\sigma_3(0)$ reveal the existence of stresses other than those of thermomechanical origin. It could originate from an asymmetry in the assembly and/or possible intrinsic stresses coming from the coating process, for example, and amplified by the cooling.

In any case, such stresses inside the $1$~mm thick quartz substrate raises the question of the corresponding stresses in the $0.35$~nm thick graphene film. Indeed, the integral $S$ of the stress through the thickness of quartz substrate, i. e. the force per unit width~\cite{EernisseJAP1972} $S=\int_{-t_q/2}^{+t_q/2} \sigma_i(0, x_2, 0) \, \mathrm{d}x_2$ is just $b_i t_q$ when stresses behave as $\sigma_i(x_2)=ax_2 + b_i$, and should be such that $|S|=\bar{\sigma_g}t_g$, where $\bar{\sigma_g}$ is the average stress in the graphene film, in a free-expansion/compression composite graphene-on-quartz device. The resulting mean value $|\bar{\sigma_g}|=b_i\frac{t_q}{t_g}$ could then be much higher than the tensile limit. To our knowledge there is no reported value of graphene intrinsic tensile strength at $4$~K but, as an indication, an intrinsic tensile strength of $130$~GPa is reported for a suspended graphene membrane at RT~\cite{Tsoukleri, LeeScience}. For metals, yield strengths at cryogenic temperatures are typically greater than that at RT~\cite{TamarinAtlas}(but what about a graphene monolayer?).  

Beyond these figures, modeling, including FEM simulation even consistent with the analog model, reaches its limits here, mainly due to the poor knowledge of the physical constants of materials at low temperatures (e.g. graphene CTE, graphene elastic constants).

Nevertheless, although the identified solutions are still affected by a rather important uncertainty, it is demonstrated that both effects, mass loading and thermomechanical effects, look well entangled and can explain the experimental frequency shifts, including positive ones. Accordingly, the measured mechanical losses are those of a stressed system.

\section{Conclusion}
As expected, coating an acoustic cavity results in a quality-factor change and a frequency shift of all the overtones of each of its eigen modes. Because of, first, its ability to be piezoelectrically excited on all its 3 thickness modes by a lateral electric field, second, each of these 3 modes reacting differently to stresses, a BAW SC-cut quartz-crystal resonator becomes attractive as a stress sensor beyond the usual mass sensor. Consequently, provided that thermomechanical stresses dominate, such a BAW SC-cut quartz-crystal resonator potentially would offer the opportunity to test/verify mechanical and thermal properties of the coating - data such as Young modulus, Poisson coefficient, CTE - even in unusual conditions, i.e. at liquid-helium temperature in our case.
 
The device tested in this study, not initially designed for use as a sensor, nevertheless demonstrates that stress effects cannot be neglected compared to those of the mass-loading in the case of a graphene single layer in contrast with usual "thin" films as Au and Cr coatings. In the present state, Q-factor measurements of this stressed resonator lead to a probable overestimation of the mechanical losses in the graphene monolayer (under a stress field), estimated at best at $8\:10^{-4}$ at $4$~K, but they could depend on the substrate nature. They also reveal a degradation of the surface roughness of the acoustic cavity by a factor of 7, because of the graphene coating, cause of diffraction and thus an increase of the losses. 

Many questions remain unanswered and improvements could be made to such a sensing system for further measurements. It would be preferable to work with a thinner resonator to limit the stress in the deposited film. The symmetry of the device should be improved by ensuring the centering of the graphene sheet to limit possible spurious modes and it would be desirable to deposit graphene on both sides despite its complexity of implementation.

\section*{Acknowledgments}
Thanks to Val\'{e}rie Soumann, from the FEMTO-ST Institute, France, for metal coatings and microscope imaging. This work was supported by Conseil R\'{e}gional de Bourgogne Franche-Comt\'{e} (France), and by the ANR-PIA (France) through the FIRST-TF Network under Grant ANR-10-LABX-48-01, the Oscillator Instability Measurement Platform under Grant ANR-11-EQPX-0033-OSC-IMP, the EUR-EIPHI Graduate School under Grant ANR-17-EURE-00002. J.B. is thankful to MESRI France, for his grant. Thanks also to the French embassy in Australia for its financial support through the Scientific Mobility Program. 
MG and MT were supported by the Australian Research Council (Australia) Grant No. DP190100071 and CE170100009.

%
\appendix
\section{Mass loading}
\label{sec:appendixmassloading}
%
\renewcommand{\theequation}{A.\arabic{equation}}
%
To quantify the frequency shifts resulting from a mass loading, let us consider the example of an infinite quartz plate of thickness $t_q$ whose normal axis is $y$ (subscript $2$ in the following equations) with the origin ($y=0$) in the centre of the thickness of the plate. The plate is infinitely coated on both sides with coating thickness $t_l$ of a material of density $\rho_l$. The boundary conditions at plate surfaces involve surface stresses $\sigma$: $\sigma_{2i}(y=+t_q/2)=- \rho_l t_l \ddot u_i(t_q/2)$ and $\sigma_{2i}(y=-t_q/2)=+ \rho_l t_l \ddot u_i(-t_q/2)$, where $i=1,2,3$, for the C, A, and B modes respectively, $\ddot u_i$ the second time-derivative of the displacement. In this case resonant frequencies of thickness modes are given by:
\begin{equation}
f_n \approx \frac{n}{2 t_q} \sqrt{\frac{\bar{c}_{2i2i}}{\rho_q}}\left[ 1-\frac{4k_{222i}^2}{n^2 \pi^2} - R  \right]
\label{eq:freq with mass loading}
\end{equation}
corresponding to a fractional frequency shift:
\begin{equation}
\frac{f_n-f_{n0}}{f_{n0}}=\frac{\Delta f_n}{f_{n0}} \approx - (1+\frac{4k_{222i}^2}{n^2 \pi^2})R,
\label{eq:fractionalfreq with mass loading and piezo}
\end{equation} 
where $f_{n0}$ denotes a frequency before coating, the odd integer $n$ denotes the OT order, $\bar{c}_{2i2i}$ is an elastic coefficient modified by piezoelectricity (pointed out by the upper bar: $\bar{c}_{2nr2}={c}_{2nr2}+\frac{e_{22n}e_{22r}}{\epsilon_{22}}$, with $e_{22i}$: piezoelectric coefficients, $\epsilon_{22}$: electric permittivity), $k_{22 \, 2i}$ is the electromechanical coupling factor, and $R$ is the ratio of the additive mass over the quartz mass i.e. $R=\frac{2 \rho_l t_l}{\rho_q t_q}$ in the case of a quartz substrate coated with layers on both faces. 
Quartz is just lightly piezoelectric, so that for the SC cut at room temperature, the quantity $k_{222i}^2=\frac{e_{22i}^2}{\epsilon_{22}\bar{c}_{2i2i}}$ can be estimated as  $1.76\times 10^{-3}$, $2.18\times 10^{-3}$, $0.46\times 10^{-3}$ for the A, B and C modes respectively. Thus, the vibration is often seen from a pure mechanical point of view for which a simplified resonance frequency shift just reads:
\begin{equation}
\frac{f_n^{'}-f_{n0}^{'}}{f_{n0}^{'}}=\frac{\Delta f_n^{'}}{f_{n0}^{'}} \approx -R,
\label{eq:fractionalfreq with mass loading}
\end{equation} 
and does not depend anymore on the vibrating mode type, A, B or C. This estimation approach has become popular in the QCM community, and known as Sauerbrey's formula~\cite{Sauerbrey, Lostis, Lu, Mecea, Reed, Johannsmann,JohannsmannBook}.
\begin{table*} [t!]
\caption{\label{tab:Dataat300K} Material parameters at $300$~K (Quartz~\cite{Bechmanncoef21958}, Au~\cite{Wang, Merle}, Cr~\cite{Merle}, Graphene~\cite{Blakslee, Bunch}). For Quartz, $C_A, C_B$, and $C_C$ are the SC-cut effective stiffness coefficients of A, B and C modes respectively.}
	\begin{tabular*}{1\textwidth}{l l l l l l}
		\hline
	Material & Density& Quartz SC-cut & Young mod. & Poisson coef.  & Shear mod.\\
	&$\rho$ (kg/m$^{3}$)& $C_{ij}$ (GPa) @ RT&$Y$ (GPa)&$\nu$&$G=\frac{Y}{2(1+\nu)}$ (GPa) \\
	\hline
	Quartz                 & $2648$ & $C_{11}=86.7$, $C_{13}=16.8$ &          &           & $(C_A \approx 121)$ \\
	&        & $C_{33}=109.9$, $C_{35}=13.0$&           &           & $(C_B \approx 41.5)$ \\
	&        & $C_{51}=-13.64$, $C_{55}=58.7$&           &           & $(C_C \approx 34.5)$ \\
	Au           & $19300$&          & $75$      & $0.44$    &   $26$ \\
	Cr                 & $7140$ &          & $275$     & $0.21$   &   $115$ \\
	Graphene & $2200$ &          & $1000$    & $0.16$   &   $430$ \\
	\hline
	\end{tabular*}

\end{table*}
\section{Effect of static stresses}
\label{sec:appendixstaticStresses}

\renewcommand{\theequation}{B.\arabic{equation}}

Effects of a static mechanical bias on elastic waves, i.e. small dynamic fields superimposed on a static bias, were intensively studied in 70's-80's~\cite{Tiersten1978, Baumhauer1973, Sinha1982, Tiersten1981} after Thurston and Brugger works in 1964 ~\cite{ThurstonPhysRev1964}. In this work, we employ Sinha-Tiersten's perturbation analysis limited to the perturbation of the elastic constants and not including dielectric or piezoelectric constant changes for example which can be justified by the weak piezoelectric coupling of quartz~\cite{Stevens1997}). In accordance to this approach, the fractional frequency change, at frequency $f=\frac{\omega}{2\pi}$, induced by a bias can be expressed as, for a pure thermoelastic problem:
\begin{equation}
\Delta \omega = \frac{1}{2\omega}\frac{ \int \int\limits_{V} \int \hat{C}_{k \alpha l \gamma} u_{\alpha,k} u_{\gamma,l} \, \mathrm{d}V }{\int \int\limits_{V} \int \rho_{0} u_{\alpha} u_{\alpha} \, \mathrm{d}V },
\label{eq:df_over_f_stress}
\end{equation}
with
\begin{eqnarray}
\hat{C}_{k \alpha l \gamma} &=& c_{k \alpha l n} w_{\gamma,n} +  c_{kml\gamma} w_{\alpha,m} \nonumber \\ &+& c_{k\alpha l\gamma a b} w_{a,b} + c_{klab} w_{a,b} \delta_{\alpha \gamma}\nonumber\\
&+& \frac{dc_{k \alpha l \gamma}}{dT}(T-T_0),
\label{eq:c_Equation_vs_w}
\end{eqnarray}
where $c_{ijkl}$ and $c_{ijklmn}$ are the second and third order elastic stiffness coefficients respectively, $w_{i,j}$ the bias displacement gradients, $u_i$ the vibration displacements, at RT~\cite{Sinha1979, Stevens1983, Ballandras2006}, within the volume $V$. The last term takes into account the fact that constants depend on temperature $T$, which is assumed to be homogeneous ($T_0$ being the reference temperature). The expression is limited to the first order derivatives of stiffness coefficients since temperature changes should also be small. It should also be mentioned that in a real BAW cavity, the active part of the resonator is anchored to its supporting rim by means of four quartz bridges. 
As a result, the thermal contraction of the crystal resonator is not strictly free but rather constrained by these bridges. 

Although Eq.~(\ref{eq:df_over_f_stress}) is usually applied at RT, it can also be used for the graphene induced stress at cryogenic temperatures. In this case, the resonator without graphene is used as a reference state assuming it is stress-free at $4$K. So, an infinite flat plate vibrating at $f_n=\frac{n}{2t}\sqrt{\frac{\bar{c}}{\rho}}$, and cooled down from RT to $4$K would exhibit a fractional frequency change:
\begin{equation} 
\frac{f_{4K}-f_{300K}}{f_{300K}}=\frac{\sqrt{1+(\alpha_1+\alpha_2+\alpha_3)\delta T}}{1+\alpha_2\delta T}\sqrt{\frac{\bar{c}_{4K}}{\bar{c}_{300K}}}.
\label{eq:deltafoverf_4K_300K}
\end{equation}
The calculation is performed by using numerical values for piezoelectric and stiffness coefficients at RT from Ref~\cite{Bechmanncoef21958}, and for the same coefficients at $4$K from Ref~\cite{Tarumi2007}. The corresponding ICTEs are calculated for the doubly-rotated quartz cut from values in Ref~\cite{Barron}, giving $\alpha_1\delta T =-2.54 \times10^{-3}$, $\alpha_2\delta T = -2.124\times10^{-3}$ and $\alpha_3\delta T = -1.65\times10^{-3}$ for the temperature change from $300$K to $4$K. This calculation gives realistic fractional frequency changes from RT to $4$K: indeed, the effective elastic constants $\bar{c}_{4K}$ have to be adjusted by less than $2\%$ to match the  experimental results, i.e. a fractional frequency change of $+14.75 \times 10^{-3}$ for the A mode, $+5.13 \times 10^{-3}$ for the B-mode, $-1.37 \times 10^{-3}$ for the C-mode when cooling down the device from RT to 4K. Such a result should be seen as an evidence for the validation of the ICTE assessments.
Moreover, it may also be reminded that temperature coefficients of various parameters are lower than $10^{-8}$ for temperatures close to $4$K: consequently, the temperature accuracy is not so critical.

The perturbation tensor $\hat{C}_{k \alpha l \gamma}$ can be expressed in terms of strains $E_{ij}$ by means of symmetry or antisymmetry properties of tensors as:
\begin{eqnarray}
\hat{C}_{k \alpha l \gamma} &=& c_{k \alpha l n} E_{n \gamma} +  c_{k m l\gamma} E_{m \alpha} \nonumber \\ &+& c_{k\alpha l\gamma a b} E_{ab} + c_{klab} E_{ab} \delta_{\alpha \gamma}\nonumber\\
&+& \frac{dc_{k \alpha l \gamma}}{dT}(T-T_0),
\label{eq:c_Equation_vs_w}
\end{eqnarray}
Stresses and strains are related by the following linear (first order) thermoelastic constitutive equations as (for convenience, the abbreviated notation, or Voigt notation, is used as follows: a pair of indices like $ij$ is replaced with a single index according to $11 \rightarrow 1$, $22 \rightarrow 2$, $33 \rightarrow 3$, $23 \rightarrow 4$, $13 \rightarrow 5$, $21 \rightarrow 6$):
\begin{equation}
\sigma_i = c_{ij}[E_j-\alpha_j \delta T] = c_{ij} E_j^\sigma
\label{eq:stress_vs_strain}
\end{equation}
or in terms of strains. Introducing compliance coefficients $s_{ij}$, the following relation can be written:
\begin{equation}
E_j = s_{ji} \sigma_i + \alpha_j \delta T =E_j^\sigma + E_j^T,
\label{eq:complince_times_stress_vs_strain}
\end{equation}
where $E_j^\sigma = s_{ji} \sigma_i$ is the stress-induced part of $E_j$ caused by external loads and displacements and/or non-uniformities in temperature or expansion properties, and $E_j^T=\alpha_j \delta T =\alpha_j(T) (T-T_0)$ refers to strains caused by free thermal expansion for a given temperature change $\delta T$ replaced with its ICTE. The perturbation tensor can be calculated from a  reference state at $4$K by means of the set of parameter values from Ref~\cite{Tarumi2007}, taking the third order elastic stiffness, unknown at $4$K, from their values at RT~\cite{Bechmanncoef21958,Thurstoncoef31966}.
Thus, the perturbation tensor is limited to a thermomechanical part and can be written:
\begin{eqnarray}
\hat{C}_{k \alpha l \gamma} &=& c_{k \alpha l n} s_{n \gamma m e} \sigma_{m e} +  c_{k m l\gamma} s_{m \alpha n e} \sigma_{n e}\nonumber \\ &+& c_{k\alpha l\gamma a b} s_{abcd} \sigma_{c d} + c_{klab}s_{abcd} \sigma_{c d} \delta_{\alpha \gamma}\nonumber \\
&=& [c_{k \alpha l n} s_{n \gamma m e}  +  c_{k a l\gamma} s_{a \alpha m e} \nonumber \\ &+& c_{k\alpha l\gamma a b} s_{abme} + \delta_{k m}\delta_{l e} \delta_{\alpha \gamma}] \sigma_{m e}.
\label{eq:csigma_vs_T}
\end{eqnarray}

For the case of acoustic waves propagating along the thickness $y$-axis, or $x_2$, in a flat resonator (no change along $x_1$ and $x_3$), the dynamic displacement gradients can be written:
\begin{equation}
u_{i,2}=\frac{\omega}{v}V_i \cos\Big[\frac{\omega}{v}x_2\Big]\sin(\omega t),
\end{equation}
with $\frac{\omega}{v}=\frac{n\pi}{t_q}$, $n$ is the OT number, $v$ the propagation speed and $V_i$ the eigenvector of the mode of interest (normalised as $V_iV_i=1$). In addition, volume integrals in Eq.~(\ref{eq:df_over_f_stress}) can be reduced to integrals over the thickness at the center, where the wave amplitude is maximum due to trapping. Thus, the stress-dependent part of the frequency shift becomes:	
\begin{eqnarray}
\Delta \omega& \simeq & \frac{1}{2\omega}\frac{ \int_{-t_q/2}^{+t_q/2} K_{me}\sigma_{me}(0, x_2, 0)\frac{\omega^2}{v^2} \cos^2(\frac{\omega}{v}x_2) \, \mathrm{d}x_2 }{\int_{-t_q/2}^{+t_q/2} \rho_{q} V_{\alpha} V_{\alpha}\sin^2(\frac{\omega}{v}x_2) \, \mathrm{d}x_2 } \nonumber \\
& \simeq &  \frac{\omega}{2v^2}\frac{\int_{-t_q/2}^{+t_q/2} K_{me}\sigma_{me}(0, x_2, 0) \cos^2(\frac{n \pi}{t_q} x_2) \, \mathrm{d}x_2 }{t_q \rho_{q}/2} 
\label{eq:dfoverfcalculated}
\end{eqnarray}
with 
\begin{equation}
K_{me} = 2c_{2\alpha 2n}s_{n\gamma me}V_{\alpha}V_{\gamma} +c_{2\alpha 2\gamma ab}s_{abme}V_{\alpha}V_{\gamma} + \delta_{2m}\delta_{2e}.
\label{eq:Ki}
\end{equation}
This relationship is applied to coated and uncoated cases in the following discussions.

\subsection{Uncoated resonator}

The uncoated resonator can be seen as a circular plate subject to extra diametrically applied forces $F$ coming from constrained contractions of its four bridges induced by cooling from RT to $4$K. A diametrical compression induces constant stresses $\sigma_i$ at the center of the quartz plate leading to a frequency shift:
\begin{eqnarray} 
\frac{\Delta \omega}{\omega}  & \simeq & \frac{1}{t_q \rho_q v^2} \int_{-t_q/2}^{+t_q/2} K_i \sigma_i \cos^2\Big[\frac{\omega}{v} x_2\Big] \, \mathrm{d}x_2 \nonumber\\
& \simeq & \frac{1}{t_q \rho_q v^2} \frac{\sigma_i t_q K_i}{2} 
\simeq \frac{\sigma_i}{2} R_i
\label{eq:dfoverfcalculated2}
\end{eqnarray}
where $i = 1, 3 ,5$, $R_i=\frac{K_i}{\rho v^2}$ are Ratajski coefficients~\cite{Ratajski}. The eigenvalue $\rho_q v^2$ and values of constants $K_i$ depend on the mode. Calculated values of these coefficients for the doubly-rotated SC-cut at $4$K are given in Table~\ref{tab:DataSCcut5K}. Stresses at the center of a circular plate with radius $r_q$ can be adapted from Ref.~\cite{Janiaud1978} for a four-point mounting with bridge in the $x_1-x_3$ plane to give:
\begin{equation}
\sigma_1\simeq \sigma_3 \simeq \frac{-2F}{\pi t_q r_q}
\label{eq:stresses_naked_resonator}
\end{equation}
while $\sigma_5\simeq0$ for a SC-cut. The diametrically applied force $F$ can be calculated by stating that a diameter change $2\Delta r_q$ of the circular plate due to free thermal expansion/contraction from $300$K to $4$K is constrained by an equivalent change in bridge length $2\Delta l$ caused by some force $F$. In free thermal expansion/contraction the diameter change along $x_1$ is $2\Delta r_q = 2r_q\alpha_1\delta T$ while a bridge along $x_1$, seen as a beam with a rectangular section $b\times t_q$ subjected to an axial force $F_1=b\: t_q \sigma_1^b$ at one end and clamped into place on the other end (this is an assumption at the rim), exhibits a length change $\frac{\Delta l}{l}=s_{1i}\sigma_i^b+\alpha_1\delta T$. Thus, stating that $\Delta r_q + \Delta l=0$ along the $x_1$ axis (the same approach is applied to the $x_3$ axis) gives:
\begin{equation}
F_1=\frac{b\: t_q}{s_{11}}\Big(\frac{r_q}{l}+1\Big)\alpha_1\delta T,
\label{eq:force_naked_resonator}
\end{equation}
where $\alpha_1\delta T$ denotes an ICTE. Consequently, assuming that the resonator rim is clamped, the naked device would exhibit a fractional frequency change $\frac{\Delta f_{4K}}{f}$ from the ideal reference state at $4$K (see Eq.~(\ref{eq:dfoverfcalculated2})) of $-0.32\times10^{-3}$ for the A-mode,$-0.18\times10^{-3}$ for the B-mode, and $+1.86\times10^{-5}$ for the C-mode.  \\

\begin{table} [t!]
	\caption{\label{tab:DataSCcut5K} Calculated Ratajski coefficients for the quartz SC-cut at $4$K ($\rho_q^{4K}=2665$kg/m$^{3}$) for the uncoated case. Second order elastic stiffness coefficients have been taken at $5$K, from Ref.~\cite{Tarumi2007}, but third order coefficients are still those from Refs.~\cite{Bechmanncoef21958,Thurstoncoef31966} at RT, because such data are not available at low temperature.}
			\begin{tabular*}{1\textwidth}{cccccc}
			\hline
			Mode & Eigenvector & Speed & $R_1$ & $R_3$ & $R_5$  \\
			   & $V_1$, $V_2$, $V_3$ &  (m/s) & \multicolumn{3}{c} {($10^{-11}$m$^2/$N)} \\
			\hline
			 A  & $0.221, 0.968, 0.119$ & $6782$ &  $-2.267$  & $1.146$ & $-2.666$  \\
			 B  & $0.211, 0.0717,0.975$ & $3939$ & $0.025$ & $-1.061$ & $1.533$ \\
			 C  & $0.952, 0.240, 0.189$ & $3580$ & $-0.101$ & $0.237$ & $1.716$ \\
			\hline
			\end{tabular*}
\end{table}

\subsection{Coated resonator}

As mentioned above, the graphene layer is deposited on one side of the quartz substrate at room temperature, and then this initially (seemingly) stress-free hybrid device is cooled down at $4$K.
Consequently, the mismatch in thermal expansion coefficients of these materials results in stresses and bending. This is true for a free expansion/contraction system and such induced stresses have to be added to stresses coming from the bridges. Free-expansion induced stresses at the center of the coated plate can be simplified as linear functions of the thickness coordinate $x_2$ (See for example ref.~\cite{Janiaud1978}, and this is also confirmed by FEM simulations), written $\sigma_i(0, x_2, 0)=a_ix_2+b_i$. As a consequence, Eq. ~(\ref{eq:dfoverfcalculated}) becomes:
\begin{eqnarray}
\frac{\Delta \omega}{\omega}  & \simeq & \frac{1}{t_q \rho_q v^2} \int_{-t_q/2}^{+t_q/2} K_i \sigma_i(0, x_2, 0) \cos^2\Big[\frac{\omega}{v} x_2\Big] \, \mathrm{d}x_2 \nonumber\\
& \simeq & \frac{1}{t_q \rho_q v^2} \frac{b_i t_q K_i}{2} 
\simeq \frac{b_i}{2} R_i, 
\label{eq:dfoverfcalculated2graphene}
\end{eqnarray}
for $i = 1, 3 ,5$.

\textbf{Simplified isotropic model.} Considering a simplified model of a quartz substrate as an isotropic material with a thin coating layer ($t_g \ll t_q$), both at homogeneous temperature $T$ with no rigid rotation around the center of the plate~\cite{Sinha1982}, thermoelastic stresses due to mismatch of both ICTEs when cooling from $T = T_0=300$~K down to $T = 4$~K can be estimated as follows. Solving this bilayer plate as an axisymmetric problem, thermoelastic stresses gives $\sigma_1=\sigma_3$ and $\sigma_2=\sigma_4=\sigma_5=\sigma_6=0$. Without any external force in free expansion/contraction conditions, and assuming in-plane strains $E$ are the same in the substrate and in the coating, the force (and moment) equilibrium are:  
\begin{equation}
\sigma_1=\sigma_3=\frac{N_q}{t_q}\Big(1-\frac{6x_2}{t_q}\Big),
\label{eq:stressesT11_T33}
\end{equation}
where the in-plane force $N_{1q} = N_{3q}=N_q$ acting in quartz is related to that in the graphene coating $N_g$ based on the relationship
\begin{equation}
N_q+N_g=\frac{Y_g t_g}{1-\nu_g}(E-\alpha_g\delta T)+\frac{Y_q t_q}{1-\nu_q}(E-\alpha_q\delta T)=0.
\label{eq:strain_graphene_quartz}
\end{equation}
From this equation involved forces can be simplified as:
\begin{eqnarray}
N_q=-N_g &=&\frac{\frac{Y_q t_q}{1-\nu_q}\frac{Y_g t_g}{1-\nu_g}}{\frac{Y_q t_q}{1-\nu_q}+\frac{Y_g t_g}{1-\nu_g}}(\alpha_g-\alpha_q)\delta T \nonumber \\
& \simeq &\frac{Y_g t_g}{1-\nu_g}(\alpha_g-\alpha_q)\delta T,
\label{eq:Normalforce}
\end{eqnarray}
because $t_g\ll t_q$, even if the graphene Young modulus is much greater than that of quartz ($Y_g \simeq  1$TPa). Following the approach discussed above, infinitesimal ${\alpha_i}\delta T$ is replaced with corresponding ICTE $\int_{T_0}^{T}\alpha_i \, \mathrm{d}T$, or, equivalently, by $\overline{\alpha}_i \Delta T$, where $\overline{\alpha}_i$  is the average of respective CTEs over $\{T_0, T\}$~\cite{Hutchinson}. 

The effect of the four-bridge clamping is taken into account like in the case of a uncoated quartz. It is argued that $\Delta r_q+\Delta l=0$ along bridge axis very close to $x_1$ and $x_3$ and $\frac{\Delta r_q}{r_q}=E$ for the strain $E$ extracted from Eq.~(\ref{eq:strain_graphene_quartz}):
\begin{equation}
E\simeq\frac{Y_g t_g}{Y_q t_q}\frac{1-\nu_q}{1-\nu_g}\alpha_g\delta T + \alpha_q\delta T.
\end{equation}
Here, the first term of the right-hand side of the equation can be identify as an excess strain $\Delta E$ in comparison with the strain $E\simeq\alpha_q\delta T$ of an uncoated disk of quartz in free expansion/contraction. Consequently the corresponding applied diametrical force due to bridge clamping along $x_1$ (and similarly along $x_3$) becomes:
\begin{equation}
F_1=\frac{Y_g t_g}{Y_q t_q}\frac{(1-\nu_q)}{(1-\nu_g)}\frac{b\: t_q}{s_{11}}\frac{r_q}{l}\alpha_g\delta T+\frac{b\: t_q}{s_{11}}\Big(\frac{r_q}{l}+1\Big)\alpha_1\delta T.
\label{eq:force_naked_resonator}
\end{equation}
This force is very close to the calculated one for an uncoated substrate because expansion/contraction stress effects due to the addition of the graphene layer are negligible due to $Y_g t_g\ll Y_q t_q$.

Comparing resulting frequency shifts for the case with (Eq.~(\ref{eq:dfoverfcalculated2graphene})) and without (Eq.~(\ref{eq:dfoverfcalculated2})) graphene coating, the fractional frequency difference is written as:
\begin{equation}
\frac{f_{4K\_g}-f_{4K}}{f_{4K}}\simeq \frac{R_i}{2}\frac{Y_g}{1-\nu_g}\frac{t_g}{t_q}(\alpha_g-\alpha_q)\delta T. 
\label{eq:dfoverf_after_before}
\end{equation}

\textbf{Anisotropic substrate with isotropic film.} When considering an isotropic film coated on an anisotropic substrate and assuming that expansion is free along the thickness of this bilayer plate, thermoelastic constitutive relationships relation stresses $T_i$ to strains $E_i$ can be written as:
\begin{eqnarray}
\sigma_1^f &=& B[(E_1-\alpha \delta T)+\nu(E_3-\alpha \delta T)] \nonumber \\
\sigma_3^f &=& B[(E_3-\alpha \delta T)+\nu(E_1-\alpha \delta T)] \nonumber \\
\sigma_5^f &=& 2G E_5,
\label{eq:film_stress}
\end{eqnarray}
for the graphene film, with $B=\frac{Y}{1-\nu^2}$ and $G=\frac{Y}{2(1+\nu)}$,
\begin{equation}
\sigma_i^s = c_{ij}[E_j-\alpha_j \delta T] \text{$\:\:\:i,j=1,3,5$}
\label{eq:substrate_stress}
\end{equation}
for the quartz substrate, according to the in-plane coordinate axis $x-z$ (for simplicity, i.e. $x_1 - x_3$), and $y$ (or $x_2$) along the thickness of the bilayer plate, $y=0$ being at the center of the quartz substrate.\\
Strains can be expressed in midplane strains $E_{mj}$ added to effects of midplane curvatures $\kappa_{mi}, $ as:
\begin{equation}
E_j=E_{mj}-(y-y_m)\kappa_j    \text{$\:\:\:j=1,3,5$},
\label{eq:strains}
\end{equation}
where $y_m$ denotes the midplane location~\cite{Reissner, Mindlin, Wu2008}. \\
Stresses in the quartz substrate, $\sigma_i^s$ (see Eq.~\ref{eq:substrate_stress}) can then be reached in the following way.\\  
Without any external force and moment, balance equations are ($i =1, 3, 5)$:
\begin{equation}
\int\limits_{-t_q/2}^{+t_q/2} \sigma_i^s \, \mathrm{d}y + \int\limits_{+t_q/2}^{+t_q/2+t_f} \sigma_i^f \, \mathrm{d}y=0,
\label{eq:forces}
\end{equation}
\begin{equation}
\int\limits_{-t_q/2}^{+t_q/2} \sigma_i^s(y-y_m) \, \mathrm{d}y + \int\limits_{+t_q/2}^{+t_q/2+t_f} \sigma_i^f(y-y_m) \, \mathrm{d}y=0
\label{eq:moment}
\end{equation}
Substituting Eqs.~\ref{eq:film_stress} to \ref{eq:strains} in the force balance equation, Eq.~\ref{eq:forces}, results in an expression that can be split in a first one regarding forces induced by the midplane strains ($j=1,3,5$):
\begin{equation}
\begin{split}
c_{1j}(E_{mj}&-\alpha_j \delta T)t_q \\
&+B[(E_{m1}-\alpha \delta T)+\nu(E_{m3}-\alpha \delta T)]t_f=0
\end{split}
\label{eq:midplane1}
\end{equation}
\begin{equation}
\begin{split}
c_{3j}(E_{mj}&-\alpha_j \delta T)t_q\\
&+B[(E_{m3}-\alpha \delta T)+\nu(E_{m1}-\alpha \delta T)]t_f=0
\end{split}
\label{eq:midplane3}
\end{equation}
\begin{equation}
c_{5j}(E_{mj}-\alpha_j \delta T)t_q+2GE_{m5}t_f=0,
\label{eq:midplane5}
\end{equation}
and a second one regarding forces induced by curvatures and twist ($j=1,3,5$):
\begin{eqnarray}
c_{1j}\kappa_jy_mt_q-B(\kappa_1+\nu\kappa_3)[\frac{t_q+t_f}{2}-y_m]t_f=0 \nonumber \\
c_{3j}\kappa_jy_mt_q-B(\kappa_3+\nu\kappa_1)[\frac{t_q+t_f}{2}-y_m]t_f=0 \nonumber \\
c_{5j}\kappa_jy_mt_q-2G\kappa_5[\frac{t_q+t_f}{2}-y_m]t_f=0.
\label{eq:curvature}
\end{eqnarray} 
Because of a negligible film thickness $t_f$ ($t_f << t_q$), the last set of Eqs.~\ref{eq:curvature} is approximately validated with a midplane location at $y_m \approx t_f/2 \approx 0$. Thus, with this value $y_m\approx0$, Eqs~\ref{eq:moment} describing the moment balance can be simplified as ($j=1,3,5$):
\begin{equation}
\begin{split}
&-c_{1j}\kappa_j\frac{t_q^3}{12}\\
&+B[(E_{m1}-\alpha \delta T)+\nu(E_{m3}-\alpha \delta T)]\frac{t_q+t_f}{2}t_f\\
&-B(\kappa_1+\nu\kappa_3)[(\frac{t_q}{2}+t_f)(\frac{t_q}{2})t_f+\frac{t_f^3}{3}] =0
\end{split}
\label{eq:moment_ym0_1}
\end{equation} 
\begin{equation}
\begin{split}
&-C_{3j}\kappa_j\frac{t_q^3}{12}\\
&+B[(E_{m3}-\alpha \delta T)+\nu(E_{m1}-\alpha \delta T)]\frac{t_q+t_f}{2}t_f\\
&-B(\kappa_3+\nu\kappa_1)[(\frac{t_q}{2}+t_f)(\frac{t_q}{2})t_f+\frac{t_f^3}{3}] =0
\end{split}
\label{eq:moment_ym0_3}
\end{equation} 
\begin{equation}
\begin{split}
&-C_{5j}\kappa_j\frac{t_q^3}{12}\\
&+2GE_{m5}\frac{t_q+t_f}{2}t_f\\
&-2G\kappa_5[(\frac{t_q}{2}+t_f)(\frac{t_q}{2})t_f+\frac{t_f^3}{3}] =0
\end{split}
\label{eq:moment_ym0_5}
\end{equation} 
Midplane strains $E_{mj}$, $j=1, 3, 5$, can therefore be extracted from Eqs~\ref{eq:midplane1},~\ref{eq:midplane3},~\ref{eq:midplane5}, and substituted in Eqs~\ref{eq:moment_ym0_1},~\ref{eq:moment_ym0_3},~\ref{eq:moment_ym0_5}, to get curvatures $\kappa1, \kappa3$ and twist $\kappa5$, to finally calculate the thermomechanical stresses in the quartz substrate through Eq.~\ref{eq:strains},~\ref{eq:substrate_stress}, to get the induced fractional frequency change, Eq.~\ref{eq:dfoverfcalculated2} $(i=1, 3, 5)$.
%
\section{Impurities}
\label{sec:appendiximpurities}
\begin{figure}[h!]
	\centering
	\includegraphics[width=3.25in]{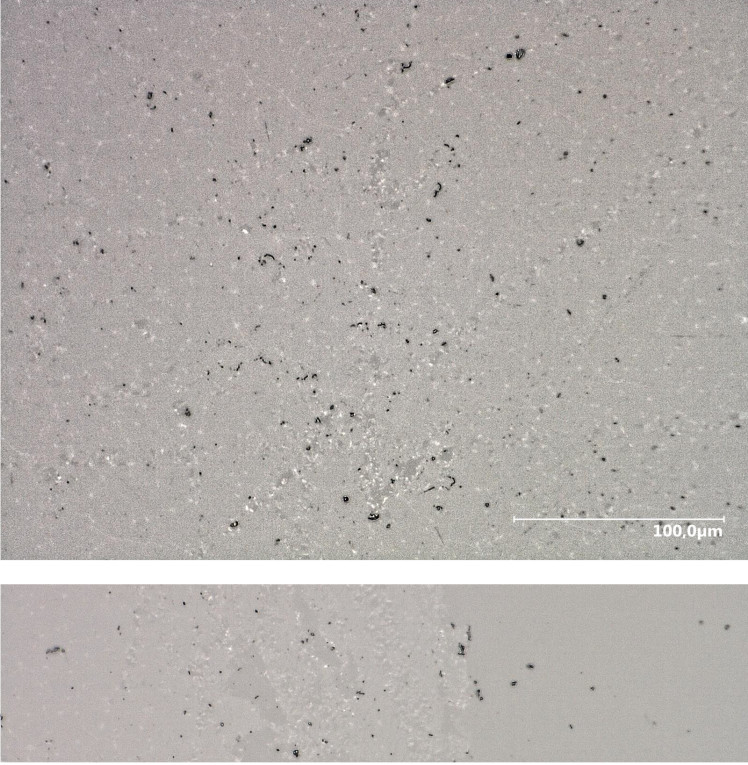}
	\caption{Impurities on the deposited graphene layer at room temperature, in the laboratory environment, once the device is removed from the cryocooler vacuum chamber. Top: close to the center. Bottom: at the edge of graphene layer.}
	\label{fig:impurities}
\end{figure}

The optical microscope image of the graphene-coated surface Fig.\ref{fig:impurities} shows existing impurities, actually dust. This photograph was taken in the environment of an ordinary laboratory room, after the device was removed from the cryogenic vacuum chamber at the end of tests at 4K. Before these tests, during its installation in the cryogenics, the device is also exposed to dusts. By counting the visible particles of $1$, $2$ and $5\:\mu m$ in the upper part (top) of Fig.\ref{fig:impurities} and assuming them spherical with an average density $1200\: kg/m^3$~\cite{Whyte2016AirbornePD}, the mass per unit area that they represent is a little less than $20\:ng/mm^2$.

Even though the estimated amount of impurities is in good agreement with frequency shifts assumed to originate from mass loading, it is impossible to say if, during the tests at 4K under vacuum, the rate of impurities varies with a decrease of the impurities collected before, at room temperature, by vacuum pumping for example, and/or with possibly a new specific contamination at 4K (adsorption, cryo-trapping effects, etc.).\\
In the lower part (bottom) of Fig.\ref{fig:impurities}, showing the edge of the graphene layer, we can note the difference in "granularity" between the graphene surface (left) and that of quartz (right) which leads to say that the roughness differs and would justify an increase of losses by wave scattering.

%

\section*{References}

%


\end{document}